\documentclass[conference]{IEEEtran}
\IEEEoverridecommandlockouts
\usepackage[normalem]{ulem} 
\usepackage{xcolor}
\usepackage[most]{tcolorbox}
\usepackage{cancel}

\usepackage{booktabs}

\usepackage{ifthen}
\newboolean{showcomments}
\setboolean{showcomments}{true} 
\ifthenelse{\boolean{showcomments}}
  {\newcommand{\nb}[2]{
    \fcolorbox{gray}{yellow}{\bfseries\sffamily\scriptsize#1}
    {\sf\small$\blacktriangleright$\textit{#2}$\blacktriangleleft$}
   }
   
  }
  {\newcommand{\nb}[2]{}
   
  }

\usepackage{cite}
\usepackage{amsmath,amssymb,amsfonts}
\usepackage{tabularray,threeparttable}
\usepackage{algorithmic}
\usepackage{graphicx}
\usepackage{subcaption}
\usepackage{xcolor,colortbl}
\usepackage{textcomp}
\usepackage{xcolor}
\usepackage{mathtools} 
\usepackage{amsmath} 
\usepackage{url,hyperref}
\def\BibTeX{{\rm B\kern-.05em{\sc i\kern-.025em b}\kern-.08em
    T\kern-.1667em\lower.7ex\hbox{E}\kern-.125emX}}

\usepackage{multirow}    
\begin{document}




\title{A data-flow oriented software architecture for heterogeneous marine data streams
}

\author{\IEEEauthorblockN{Keila Lima\IEEEauthorrefmark{1}, Ngoc-Thanh Nguyen\IEEEauthorrefmark{1}, Rogardt Heldal\IEEEauthorrefmark{1}, Lars Michael Kristensen\IEEEauthorrefmark{1}, \\Tosin Daniel Oyetoyan\IEEEauthorrefmark{1}, Patrizio Pelliccione\IEEEauthorrefmark{4}, Eric Knauss\IEEEauthorrefmark{3}}
\IEEEauthorrefmark{1}Western Norway University of Applied Sciences, Norway \\
\IEEEauthorrefmark{3}Chalmers $|$ University of Gothenburg, Sweden \\
\IEEEauthorrefmark{4}Gran Sasso Science Institute, Italy}


\maketitle

\begin{abstract}
Marine in-situ data is collected by sensors mounted on fixed or mobile systems deployed into the ocean. This type of data is crucial both for the ocean industries and public authorities, e.g., for monitoring and forecasting the state of marine ecosystems and/or climate changes. Various public organizations have collected, managed, and openly shared in-situ marine data in the past decade. Recently, initiatives like the Ocean Decade Corporate Data Group have incentivized the sharing of marine data of public interest from private companies aiding in ocean management. However, there is no clear understanding of the impact of data quality in the engineering of systems, as well as on how to manage and exploit the collected data.

In this paper, we propose
main architectural decisions and a data flow-oriented component and connector view for marine in-situ data streams. Our results are based on a longitudinal empirical software engineering process, 
and driven by knowledge extracted from the experts in the marine domain from public and private organizations, 
and challenges identified in the literature. 
The proposed software architecture is instantiated and exemplified in a prototype implementation. 

\textit{$[$The paper is accepted at the 21st IEEE International Conference on Software Architecture (ICSA 2024).$]$}

\end{abstract}



\section{Introduction}
\label{sec:intro}
Marine in-situ data is crucial for industries operating at sea in sectors and domains such as oil and gas, fish farming and aquaculture, wind farms, and shipping. 
Moreover, for public authorities, this type of data is crucial for monitoring and forecasting the state of marine ecosystems, as well as climate change. 
In these domains, marine in-situ data is collected by sensors for leak detection or other environmental monitoring such as temperature or pH, and integrity monitoring of sub-sea structures.
The infrastructures and sensor systems deployed into the ocean provide (near) real-time data streams. Due to their high associated deployment and maintenance costs, they must operate for weeks and up to years without intervention.


An emerging technological trend for the communication connectivity of ocean-deployed sensors is the Internet-of-Underwater-Things (IoUT)~\cite{ref14}. 
Data flow from IoUT to end-user applications can be decomposed into three layers: (i) data acquisition - via the underwater sensors; (ii) networking and communication - for routing data from Underwater Wireless Sensor Networks (UWSNs) to IP networks; and (iii) data management - for delivering data to end-user applications.
{These layers can also be represented from a data actor viewpoint by data producers, data service providers, and data consumer sub-systems operated by different organizations~\cite{ref13}.}

However, IoUT systems present challenges related to the marine environment, e.g., rough sea conditions, limited sensor battery capacity, and constrained communication bandwidth. More specifically, in our context, we consider non-mobile nodes and acoustics for wireless underwater communication inside the UWSN. 
Underwater acoustic communication (UAC) has low bandwidth, high propagation delay, and high bit error rate, especially with horizontal communication, as the signals are prone to distortions and can be affected by water conditions. 
Therefore, the assessment of data quality is a design concern for IoUT data-driven applications and services. The lack of understanding of data quality in systems may lead to analytics based on inadequate data by automated and AI-based systems~\cite{10188681}. This in turn may result in sub-optimal or poor decision-making, and previous studies have discussed the need to investigate the impact of data quality on the design of machine learning and AI-dependent systems~\cite{9609199,kuwajima2020engineering}.
Moreover, the European Union (EU) has made significant efforts towards regulating the usage of low-quality data in applications that could potentially impede citizens' rights. The regulatory framework proposal on artificial intelligence\footnote{\href{https://ec.europa.eu/newsroom/dae/redirection/document/75788}{https://ec.europa.eu/newsroom/dae/redirection/document/75788}} aims to ensure that high-risk AI systems adhere to these standards to establish their trustworthiness. 

The dependence of these complex and intelligent marine systems on IoUT data requires substantial knowledge on their resulting architecture. This is crucial to understand the end-to-end data quality requirements, trade-offs, and the design of the different layers, the behavior of associated components, and their interactions.  However, currently there is no clear understanding of the
impact of data quality in the engineering of systems, as well as
on how to manage and exploit the collected data.

This paper presents a data flow-oriented software architecture developed in an R\&D setting that involves stakeholders with different data-related roles within the marine domain. 
Our long-term research process results are based on the domain's specific expertise and literature, driven by data quality. Despite the domain-specific approach, data quality is a critical aspect of software systems with the increasing incorporation of data analytics tools. Thus, implications, requirements, and architecture to address them need to be studied. With this aim, we analyzed software and data quality requirements throughout the data flow, proposing a set of architectural decisions. These have been incorporated into the architecture and 
have been instantiated and exemplified on a prototype implementation. 
The ambition of this paper is to aid practitioners in the marine domain in making informed decisions when architecting their systems. 


\section{Research methodology} 
\label{sec:method}
Our study is motivated by a real-world need for a software architecture accommodating wireless smart ocean observation systems (hereafter, referred to as \texttt{System}). 
We follow the Design Science approach~\cite{hevner2010design} to conceive and design 
the architectural decisions (Section~\ref{sec:decisions}) and the software architecture for \texttt{System} (Section~\ref{sec:prototype_arch}), addressing our identified need.


\subsection{Context}
\label{sec:context}
The research context plays a crucial role in Design Science~\cite{petersen2009context}.
Our study involves 
16 marine organizations that collaborate 
to build \texttt{System}, as detailed in Table~\ref{tab:partners} (gray rows indicate software-related stakeholders). \texttt{System} accommodates representative use cases within ocean industries covering aquaculture, environmental monitoring, and energy. The stakeholders play different data-related  roles in building \texttt{System}: 

\begin{table}[!t]
  \caption{Organizations involved in building \texttt{System}
  }
  \label{tab:partners}
  {\footnotesize
  \begin{tabular}{|p{.20cm}|p{.8cm}|p{5.3cm}|p{0.9cm}|}
    
    \hline
    \textbf{ID} & \textbf{Size}     & \textbf{Profile }    & \textbf{Role}   \\
       & \textbf{(emp.)} &          &       \\
\hline
    1       & $<$250    & Producer of oceanographic sensors and supplier of sensory services & P, S, C\\
    \hline 
    \cellcolor{gray!40} 2       & \cellcolor{gray!40} $>$250    & \cellcolor{gray!40} Research institute on technology development for sensors and decision support & \cellcolor{gray!40} C\\
    \hline 
    \cellcolor{gray!40} 3       & \cellcolor{gray!40} $>$250  & \cellcolor{gray!40} Underwater communication and marine robotics manufacturer  & \cellcolor{gray!40} P, S \\ 
    \hline
    4      & $>$250  & Defence research and development institute  &  P, S \\ 
    \hline
    5     & --  & Marine industry cluster to foster technology synergies in the sector &  I \\
    \hline 
    6  & $>$250 & Oil exploration and petroleum resources development company &  C  \\ 
    \hline
    \cellcolor{gray!40} 7  & \cellcolor{gray!40} $<$250 & \cellcolor{gray!40} Oceanographic sensors manufacturer and sensor services provider &  \cellcolor{gray!40} P, S  \\ 
    \hline
    \cellcolor{gray!40} 8     & \cellcolor{gray!40} $<$250  & \cellcolor{gray!40} Manufacturer of underwater communication systems and monitoring services provider & \cellcolor{gray!40} P, S  \\
    \hline 
    \cellcolor{gray!40} 9      & \cellcolor{gray!40} $>$250  & \cellcolor{gray!40} IT and digital solutions consultancy &  \cellcolor{gray!40} S, C  \\
    \hline 
    10      & --  & Innovation industry cluster to foster national and international R\&D cooperation &  I \\
    \hline 
    11     & $<$250    & Research institute focus in marine, cryospheric and atmospheric studies &  P, S, C \\
    \hline
    12      & $>$250  & Subsea systems development, operation, and consultancy &  P, S  \\
    \hline 
    \cellcolor{gray!40} 13     & \cellcolor{gray!40} $>$250  & \cellcolor{gray!40} Marine research institute    &  \cellcolor{gray!40} P, S, C \\
    \hline 
    14      & $<$250  & Offshore telecommunications carrier &  S  \\
    \hline 
    15     & --  & Cluster of research institutes, governmental bodies, companies related to seafood industry  &  I  \\
    \hline
    16      & $<$250  & Subsea inspection systems development and operation &  P, S  \\
    \hline 
\end{tabular}
}
\vspace{-0.5cm}
\end{table}

\begin{itemize}
    \item \textbf{Data Producers} (Role \textbf{P}): organizations responsible for developing and integrating underwater sensors into Underwater Wireless Sensor Networks (UWSNs).
    \item \textbf{Data Service Providers} (Role \textbf{S)}: organizations providing services for transmitting, processing, or sharing data in underwater and onshore environments.
    \item \textbf{Data Consumers} (Role \textbf{C}): organizations consuming data produced by \texttt{System} for data analytics purposes. They provide crucial requirements for building an architecture that can be parameterized for specific use cases.
    \item \textbf{Industry Clusters} (Role \textbf{I}): 
    clusters of organizations that are responsible for knowledge and technology transfer between research and the industrial settings to support innovation of \texttt{System} but do not directly involve its development. 
\end{itemize}

\subsection{Research process}\label{sec:method_process}
\begin{figure}[h]
    \centering
    \includegraphics[trim={6.2cm 2.5cm 3cm 2cm},clip,width=\linewidth]{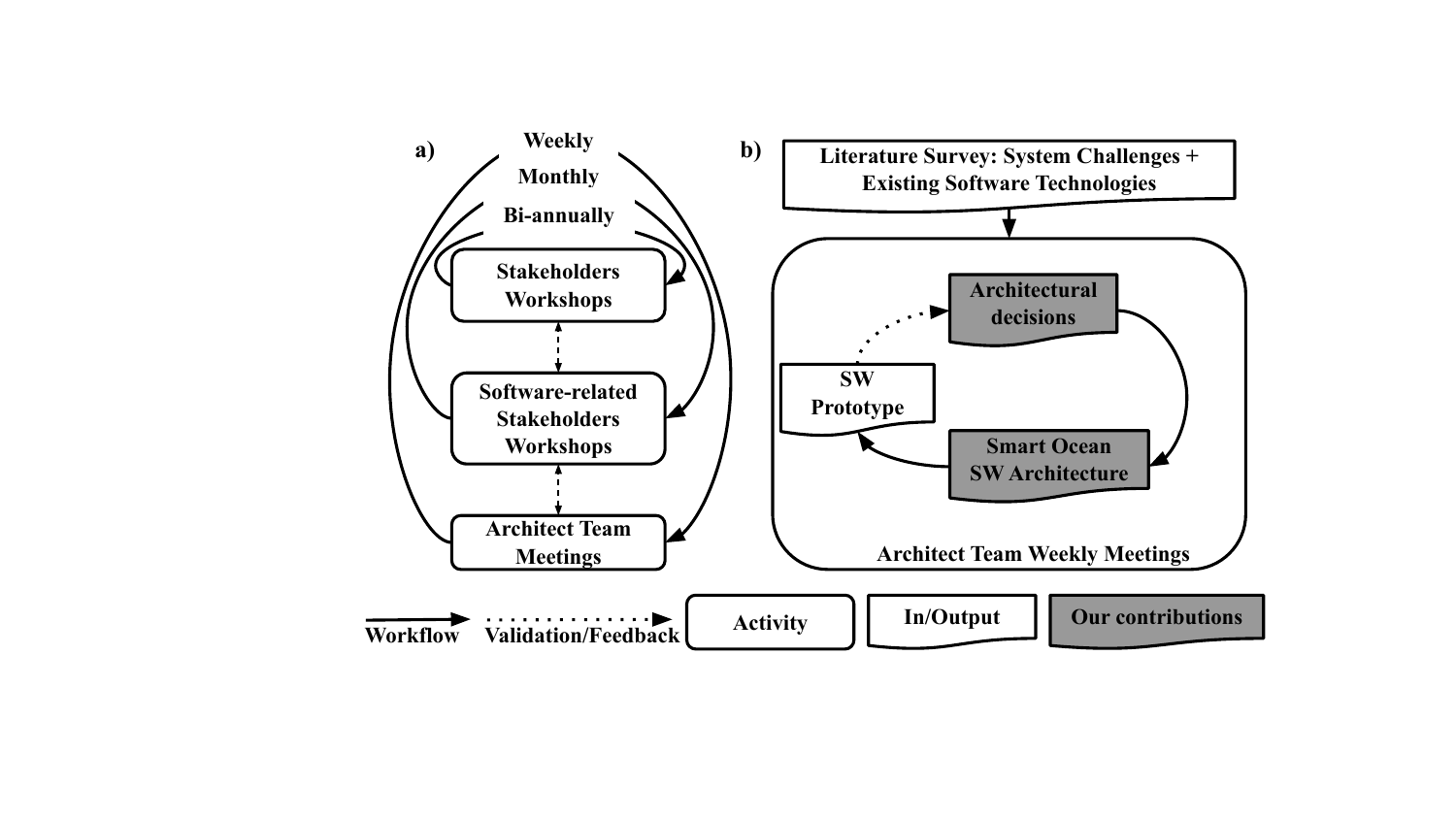}
    \vspace{-.4cm}
    \caption{Research process with: a) depicting the relevance cycle; and b) illustrating the rigour cycle and solution design process}
    \label{fig:research_process}
\end{figure}

Figure~\ref{fig:research_process} depicts the relevance, rigor, and design cycles following the Design Science research approach~\cite{hevner2010design}.

\textbf{Relevance cycle:} Figure~\ref{fig:research_process}a illustrates how we, the architect team, ensure that the produced artifacts are relevant to the needs of the stakeholders listed in Table~\ref{tab:partners}. 
The relevance cycle of this study started in January 2022. Each team member worked individually and the whole team gathered for 2 hours every week, for status updates and to discuss issues encountered while producing the artifacts. We also organized 2-hour monthly workshops with the software-related stakeholders. These workshops aimed to elicit the software requirements, determine the architecture to be developed, identify necessary architectural decisions, and validate our ongoing efforts. The artifacts were presented in 2-day bi-annual workshops bringing together all stakeholders of the project together. The primary aim of the bi-annual workshops was to ensure the alignment of all the software architecture's components of \texttt{System}, including underwater data acquisition, network communication, and software technologies. 

\textbf{Rigor and Design cycles:} Figure~\ref{fig:research_process}b shows the cycles of rigor and design. 
Our initial exploratory studies~\cite{ref14,ref13,ref15} on building \texttt{System} and the associated literature surveys served as starting point for this stage of the research process.
We took into account the challenges presented in the literature (Section~\ref{sec:challenges}), and existing software technologies (Section~\ref{sec:decisions} and~\ref{sec:prototype_arch}).
Those findings resulted in decisions and a software architecture for \texttt{System} (Sections~\ref{sec:decisions} and~\ref{sec:prototype_arch}). A prototype implementation of the software architecture was made to validate and adjust the decisions accordingly. Details of 
our previous studies and
 the software components can be found in our companion package~\cite{replication_package}.

\subsection{Threats to validity}\label{sec:validity}
\subsubsection{Construct Validity}
 Since the gathered challenges from the literature were a pivotal point in our research methodology, we do not claim to have identified all the existing challenges for data-driven systems that process marine in-situ data in realtime. We focus on the ones that our architectural decisions address in the intersection of our design cycle, the rigor cycle outcomes, and the relevance cycle inputs and validation. While the authors gained domain knowledge, gathering papers in the literature diminished until the understanding of the challenge reached a saturation point. Furthermore, to analyze the challenges, we use standardized quality models, which were extended by inserting minimal changes (two additional quality attributes) supported by the literature.

\subsubsection{Internal Validity}
To reduce the threat of misinterpretation of the identified challenges, we used views from different reference papers from literature collected over the research period. To validate the rationale behind the challenges, they were discussed several times with the stakeholders in the activities held in the relevance cycle (see Figure~\ref{fig:research_process}a).
We proposed an architecture that reflects the decisions whose rationale incorporates existing challenges from different studies, technological solutions, and stakeholders' domain expertise. 
Furthermore, we do not limit the validation of the resulting architecture to the stakeholders' interaction chain; we also integrate feedback from the prototype technical implementation process. The integration and prototyping processes only started when sufficient knowledge of the domain was acquired to ensure that as many factors as possible were taken into consideration in the solution-seeking stage.

\subsubsection{External Validity}
The research process focused on requirements for a set of heterogeneous data-driven use cases within the marine domain. Our resulting flexible and parameterized architecture is a reflection of this.
Consequently, in contexts that share similar challenges as those described throughout Section~\ref{sec:challenges}, the related architectural decisions can be suitable for similar systems. 
Lastly, the architectural decisions entail some limitations, and their usage in other contexts has to be analyzed carefully to ensure the quality of the resulting architecture.

\subsubsection{Reliability}
To improve the reliability of the study, we provide the documents from which Tables~\ref{tab:challenges} and~\ref{tab:decisiosn} were extracted. As for the resulting architecture, we also provide the initial draft upon which new knowledge has been integrated between the many iterations over our relevance, design and rigor cycles (Section~\ref{sec:method}). Because of intellectual property rights (IPR) constraints and the cross-organizational context, the source code of the end-to-end prototype implementation cannot be provided. Runtime excerpts of the system's operation is provided. All this data is available in the companion package of the study in~\cite{replication_package}.

\section{Challenges}
\label{sec:challenges}
The system behavior is highly dependent on data quality, and our aim is to analyze an architecture from the point of view of its heterogeneous data flows. We have therefore analyzed the gathered challenges using the quality attributes for software and data products from the ISO/IEC 25010~\cite{iso25010} and the ISO/IEC 25012~\cite{iso25012_2} quality models, respectively. We introduced two additional quality attributes on the quality model in order to accommodate challenges not captured by ISO/IEC 25010  and ISO/IEC 25012: Energy efficiency as suggested in~\cite{bass2003software} and Explainability from~\cite{9609199,habibullah2023non}. 

The energy efficiency quality attribute is identified across different studies~\cite{bass2003software,VENTERS2018174,koccak2015integrating} as a characteristic to be considered when evaluating  of software systems as it contributes to the software sustainability. 
Although the ISO/IEC 25010~\cite{iso25010} encompasses a resource utilization sub-characteristic under performance efficiency quality attributes, these usually refer to computational resources (e.g. CPU, storage, and network usage). This can be ambiguous if energy is perceived as a resource because it is a trade-off to the computational resources. 
Explainability has emerged as a quality attribute associated with machine learning and other AI-based components due to their ``black box'' characteristics. It is often associated with the quality of data used by such systems since its runtime behavior depends on data~\cite{9609199}, which in turn influences the decision-making process of these systems. 

Below we discuss 15 of the identified challenges from the perspective of the software and the data quality attributes. 
The challenges and their impacted software and data quality attributes of \texttt{System} are summarized in Table~\ref{tab:challenges}. 

\begin{table*}[h]
\caption{Software architecture relevant challenges on stationary wireless marine observing systems}
\label{tab:challenges}
\centering
\scriptsize
{
\centering
\begin{tabular}{|c|l|c|l|ll|l|}
\hline
\multirow{2}{*}{\textbf{ID}} & \multicolumn{1}{c|}{\multirow{2}{*}{\textbf{Challenge}}} & \multirow{2}{*}{\textbf{Layer}} & \multicolumn{1}{c|}{\multirow{2}{*}{\textbf{\begin{tabular}[c]{@{}c@{}}Quality of Software \\ Product Impacted\end{tabular}}}} & \multicolumn{2}{c|}{\textbf{Quality of Data Product Impacted}} & \multicolumn{1}{c|}{\multirow{2}{*}{\textbf{References}}} \\ \cline{5-6}
 & \multicolumn{1}{c|}{} &  & \multicolumn{1}{c|}{} & \multicolumn{1}{c|}{\textbf{Inherent}} & \multicolumn{1}{c|}{\textbf{System-dependent/Both}} & \multicolumn{1}{c|}{} \\ \hline
CH1 & Limited battery & \multirow{4}{*}{\begin{tabular}[c]{@{}c@{}}Underwater \\ Data Acquisition\end{tabular}} & Reliability (availability)& \multicolumn{1}{l|}{\begin{tabular}[c]{@{}l@{}}Accuracy,\\ Credibility,\\Completeness\end{tabular}} & Recoverability & \cite{ref2,ref3,ref5,ref11,ref13,ref14} \\ \cline{1-2} \cline{4-7} 
CH2 & Physical vulnerability &  & Reliability (availability) & \multicolumn{1}{l|}{\begin{tabular}[c]{@{}l@{}}Accuracy,\\ Credibility,\\Completeness\end{tabular}} & Recoverability & \cite{ref3,ref4,ref5,ref7,ref8,ref11,ref14} \\ \cline{1-2} \cline{4-7} 
CH3 & Lack of device output format standards &  & \begin{tabular}[c]{@{}l@{}} Compatibility \\ (Interoperability)\end{tabular} & \multicolumn{1}{l|}{\begin{tabular}[c]{@{}l@{}}Credibility,\\ Completeness,\\Consistency\end{tabular}}  & \begin{tabular}[c]{@{}l@{}} Compliance \\ Portability\end{tabular} & \begin{tabular}[c]{@{}l@{}} \cite{ref3,ref5,ref7} \\ \cite{ref14,ref15} \end{tabular} \\ \hline 
CH4 & Unreliable UAC links &  & \begin{tabular}[c]{@{}l@{}} Reliability (Fault \\Tolerance) \end{tabular} & \multicolumn{1}{l|}{\begin{tabular}[c]{@{}l@{}}Credibility,\\Completeness\end{tabular}} & Recoverability & \begin{tabular}[c]{@{}l@{}} \cite{ref2,ref4,ref5,ref7} \\ \cite{ref9,ref10,ref11,ref14} \end{tabular} \\ \cline{1-2} \cline{4-7} 
CH5 & Missing raw data & \multirow{6}{*}{\begin{tabular}[c]{@{}c@{}}Network \\ Communication\end{tabular}}  & Explainability* & \multicolumn{1}{l|}{\begin{tabular}[c]{@{}l@{}}Credibility,\\ Completeness\end{tabular}} & Understandability  & \cite{ref13,ref14} \\ \cline{1-2} \cline{4-7} 
CH6 & Environmental impact of UAC & & \begin{tabular}[c]{@{}l@{}}Performance efficiency \\ (resource utilization)\end{tabular} & \multicolumn{1}{l|}{\begin{tabular}[c]{@{}l@{}}Credibility,\\Completeness\end{tabular}} & N/A & \cite{ref1,ref5,ref14} \\ \cline{1-2} \cline{4-7} 
CH7 & \begin{tabular}[c]{@{}l@{}}Low communication bandwidth and \\ high latency of UAC and OTA links\end{tabular} &  & \begin{tabular}[c]{@{}l@{}}Performance efficiency \\ (capacity)\end{tabular} & \multicolumn{1}{l|}{\begin{tabular}[c]{@{}l@{}}Credibility, \\ Currentness\end{tabular}} & N/A & \cite{ref6,ref13,ref14,ref16} \\ \cline{1-2} \cline{4-7} 
CH8 & Security vulnerability of UAC links &  & Security & \multicolumn{1}{l|}{Credibility} & \begin{tabular}[c]{@{}l@{}}Availability, \\ Confidentiality\end{tabular} & \begin{tabular}[c]{@{}l@{}} \cite{ref3,ref6,ref11,ref12} \\ \cite{ref14,ref19} \end{tabular} \\ \cline{1-2} \cline{4-7} 
CH9 & Power-hungry UAC data transmission &  & Energy efficiency* & \multicolumn{1}{l|}{Credibility} & Recoverability & \cite{ref3,ref10,ref14} \\ \hline
CH10 & Complex data usage & \multirow{6}{*}{\begin{tabular}[c]{@{}c@{}}Data \\ Management\end{tabular}} & \begin{tabular}[c]{@{}l@{}} Usability (Appropriateness \\ recognizability)\end{tabular} & \multicolumn{1}{l|}{N/A} & Understandability & \cite{ref13,ref14,ref15} \\ \cline{1-2} \cline{4-7} 
CH11 & Lack of automatic QC procedures &  & N/A & \multicolumn{1}{c|}{\textbf{All}} & \multicolumn{1}{c|}{\textbf{All}} & \cite{ref13,ref14,ref15,ref17} \\ \cline{1-2} \cline{4-7} 
CH12 & Heterogeneous data sources & &  \begin{tabular}[c]{@{}l@{}} Compatibility \\ (Interoperability)\end{tabular} & \multicolumn{1}{l|}{\begin{tabular}[c]{@{}l@{}}Credibility,\\ Completeness,\\Consistency\end{tabular}} & \begin{tabular}[c]{@{}l@{}} Compliance \\ Portability\end{tabular} & \cite{ref13,ref15,ref18} \\ \cline{1-2} \cline{4-7} 
CH13 & \begin{tabular}[c]{@{}l@{}}Sensitive data \\ (Business and Defense critical)\end{tabular} &  & Security & \multicolumn{1}{l|}{N/A} & \begin{tabular}[c]{@{}l@{}}Availability, \\ Confidentiality\end{tabular} & \cite{ref13,ref15} \\ \cline{1-2} \cline{4-7} 
CH14 & Data reusability &  & (Re)usability & \multicolumn{1}{l|}{N/A} & \begin{tabular}[c]{@{}l@{}}Availability, \\ Confidentiality\end{tabular} & \cite{ref13,ref14,ref15} \\ \cline{1-2} \cline{4-7} 
CH15 & Access control for data sharing &  & Security & \multicolumn{1}{l|}{N/A} & \begin{tabular}[c]{@{}l@{}}Availability, \\ Confidentiality\end{tabular} & \cite{ref13,ref15} \\ \hline
\end{tabular}
}
\begin{tablenotes}
      \item \scriptsize
*  -- New adopted QA \(|\)  Abbreviations: UAC (Underwater Acoustic Communication); OTA (Over-The-Air); N/A (Not Applicable).
\end{tablenotes}
\end{table*}

\subsection{Compatibility}
\label{sec:compatibility}
\textbf{Compatibility} of \texttt{System} 
is affected in the underwater data acquisition layer because of the sensor integration. It implies that heterogeneous communication protocols and associated output data formats need to be unified at the UWSN level to ensure a common syntax and semantics within data producers' sub-systems (\textbf{CH3}). 
It affects also the data management layer as the originating dataset from the integration of measurements from different sensors is supplied by data producers' sub-systems in various formats. 
Additionally, there might be different types of data, such as time series, images, or videos, which are categorized into point-based and array-based data~\cite{ref15}. At the cloud service level, where multiple data sources are merged, interoperability is required to ensure that the different processing components interpret data based on the same metrics and criteria (\textbf{CH12}).
When data has proprietary or organization-dependent syntax and semantics, its \textbf{compliance} to standards and \textbf{portability}, i.e., ``\textit{degree to which data has attributes that enable it to be installed, replaced or moved from one system to another}''~\cite{iso25012}, can be affected.

\subsection{Explainability and Usability}
\label{sec:explainability}
Because of the low 
bandwidth and high propagation delay of UAC at the UWSN level, there is an aggregation or local processing of data to filter important information or events, which will be sent up the data delivery chain. The raw data is often kept for retransmission and can be retrieved during recovery or maintenance of the sensing platforms at sea. Despite that, storage might also be an issue when operating over a long time. Since the raw data records can be crucial for data analytics tasks, their absence can have implications in the \textbf{understandability} of data and the \textbf{explainability} of automatically detected phenomena or actions performed based on the data (\textbf{CH5}). 
Lack of metadata, which can provide information regarding the required data quality inherent dimensions hindering its adequate use for a specific context, is another challenge affecting \textbf{understandability} (\textbf{CH10}).
These challenges impact the \textbf{usability}, and more specifically the ability to ``\textit{recognize whether a product or system is appropriate for their needs}''~\cite{iso25010}. 
Also, a lack of explainability of current procedures makes it difficult to fully automate data quality control procedures (\textbf{CH11}).
Because of the high complexity of data, such activities currently require domain experts' intervention and usually cause delays of up to six months. This makes the process subjective and strongly dependent on the experience of the person in charge. 
Since marine data is prone to errors due to the uncertainty of the environment and to 
technological constraints, existing procedures often focus on assessing the inherent characteristics of data.  
Still, it is of high importance to perform data quality control before employing the data further considering the quality dimensions required for its context of use.

\subsection{Recoverability}
\label{sec:reliability}
The \textbf{Recoverability} data quality attribute, i.e., ``\textit{the degree to which data has attributes that enable it to maintain and preserve a specified level of operations and quality, even in the event of failure}''~\cite{iso25012}, is impacted by challenges affecting  \textbf{reliability}, \textbf{performance efficiency}, and \textbf{energy efficiency} software quality attributes.
Regarding \textbf{performance efficiency} (\textbf{resource utilization}), then the noise generated by the communication device's signals can negatively impact species that use acoustics to communicate or prey, e.g., by leading them to confusion or causing other changes in the behavior (\textbf{CH6}). Because of the impact of the underwater acoustic signals on marine life, there could be restrictions on the transmission windows and frequencies. This challenge has implications on the communication channels and links utilization.

Furthermore, in the network communication layer, wireless communication channels in IoUT have less available bandwidth per packet transmission (\textbf{CH7}) compared to traditional land-based channels. This is the case for the acoustics in underwater communications and in satellite over-the-air communication (OTA), where latency can vary according to atmospheric or water conditions (e.g. affected by absorption). Lastly, the OTA options (satellite and mobile networks) to transmit data into IP networks usually have higher costs associated with the carrier service. These challenges constrain the overall network \textbf{performance efficiency capacity} of IoUT software systems.

When it comes to \textbf{reliability} at the underwater data acquisition layer, the only power source for underwater equipment is the battery. Its limited resource lifetime can vary from few weeks to years depending on the data sampling rate, the precision of instruments, and the power consumption of the transmission device (\textbf{CH9}). This last factor is due to the higher consumption of energy associated with transmission devices when compared to computation done locally. Thus, \textbf{energy efficiency} is one of the main drivers of the overall system.
Additionally, at the UWSN level, underwater sensors, communication modems, and gateways are vulnerable to physical impacts. For example, equipments can be damaged by human activities (e.g., hit by ships) or by nature (e.g., dragged by sea waves, currents, or ice). This can affect sensor measurements, device integrity and their resulting data quality.

In summary, while the physical vulnerability (\textbf{CH2}) of the hardware and the limited battery (\textbf{CH1}) affects the availability of the software system (its operation), the unreliability associated with the wireless underwater network links (\textbf{CH4}) impacts the capacity of the software system to operate in the presence of faults in the communication channel. 

\subsection{Security}
\label{sec:security}
When it comes to \textbf{security} at the network communication layer, UAC is susceptible to a range of attacks (\textbf{CH8}) including packet capture, routing attacks, and blockage of a link between nodes (DoS) due to the accessibility to the wireless channel. 
Authentication and detection strategies can be implemented to mitigate the open UAC communication channel and spoofing of sensor nodes. However, implementing land-based network security methods in this context is not trivial due to the communication overhead underwater, especially when there are many nodes in the network.
A detailed description of the types of attacks and their mitigation can be found in~\cite{ref19}.  At the data management layer, two interrelated challenges impact security in terms of confidentiality and rightful access to data being shared: (i) the access to data being collected needs to be carefully filtered in terms of its category (\textbf{CH15}); and (ii) data categorization and filtering are motivated by data sharing beyond the initial context in which it was acquired (\textbf{CH14}). Some reasons behind this are regulations, better forecast and safety measures on organizations' operations areas, or adhering to initiatives to provide data for climate and marine sciences for better public perception.
In summary, the data to be shared could be categorized into (\textbf{CH13}) legally restricted, business-critical, or open-access data. An example of legally restricted data is data that is critical for the maritime territory defence because it details the bathymetry of a specific region at the sea, or stock market data. Open-access data refers to ``non-footprint'' data valuable to a third-party entity or to new applications, which requires resources to be shared.
In terms of data-related quality attributes, these challenges impacts data \textbf{availability}, i.e. ``\textit{the degree to which data has attributes that ensure that it is only accessible and interpretable by authorized users}''~\cite{iso25012}; and \textbf{confidentiality}, which is defined as ``\textit{The degree to which data has attributes that enable it to be retrieved by authorized users and/or applications}''~\cite{iso25012}.

\subsection{Data Product Inherent Quality Attributes Impacted}
\label{sec:intrinsics_QA}
\paragraph{Accuracy}
This quality attribute is impacted by all challenges that affect the sensor measurements, deviating the measurements from the observed phenomena aimed to be captured. For example, bio-fouling concerns microorganisms growing around underwater sensors. It is an unavoidable problem for artificial objects operating in the ocean. When a sensor is subject to this problem, it may produce \textbf{inaccurate} measurements, e.g., due to data drift. Another example of factors impacting data accuracy is the sensor power supply. If a battery is running low and there are inconsistencies to the sensor energy supply, there might be errors introduced in the sensors' readings, which in turn originates erroneous data. Lastly, the physical vulnerability also affects the positioning of the sensors and thus can have a negative implication on the reading. For instance, sensors deployed close to the water surface may accidentally make a reading outside of the water. Another example concerns sensors that use their static configured position as part of the data acquisition and are dragged from their initially deployed location.

\paragraph{Completeness, Consistency and Currentness}
These quality attributes are impacted by all challenges affecting the reliability and performance (in terms of network throughput) quality attributes of the data-producing system. 
The reason is that these challenges entangle the availability of software components and data transmission. The compatibility-related challenges can affect the \textbf{completeness} of the data products because of the lack of common fields and metadata available from the heterogeneous sources of data at the data acquisition and management layers. In addition to the compatibility challenges,  \textbf{consistency} with other data might be affected as well because of the correlation of measurements.
Finally, the high latency associated with the underwater and OTA communication channels greatly impacts the arrival time of data to data consumer applications. This in turn affects the \textbf{currentness} data quality attribute.

\paragraph{Credibility}
\textbf{Credibility} is a cross-cutting concern inherent data quality attribute, and it relates to ``\textit{truthfulness of origins}''~\cite{iso25012}. In the IoUT context, it is not possible to have the ground truth for all measurements and most quality control procedures focus on investigating the closeness of the data to its true value~\cite{ref17}. 
Trust is also impacting the cross-organizational level of the system because of all the uncertainty affecting the data streams until its (re)use~\cite{ref13}.

\section{Architectural Decisions}
\label{sec:decisions}
 
\begin{figure*}[htb!]
    \centering
    \includegraphics[trim={0cm 0.25cm 0.5cm 0.25cm},clip,width=0.8\linewidth]{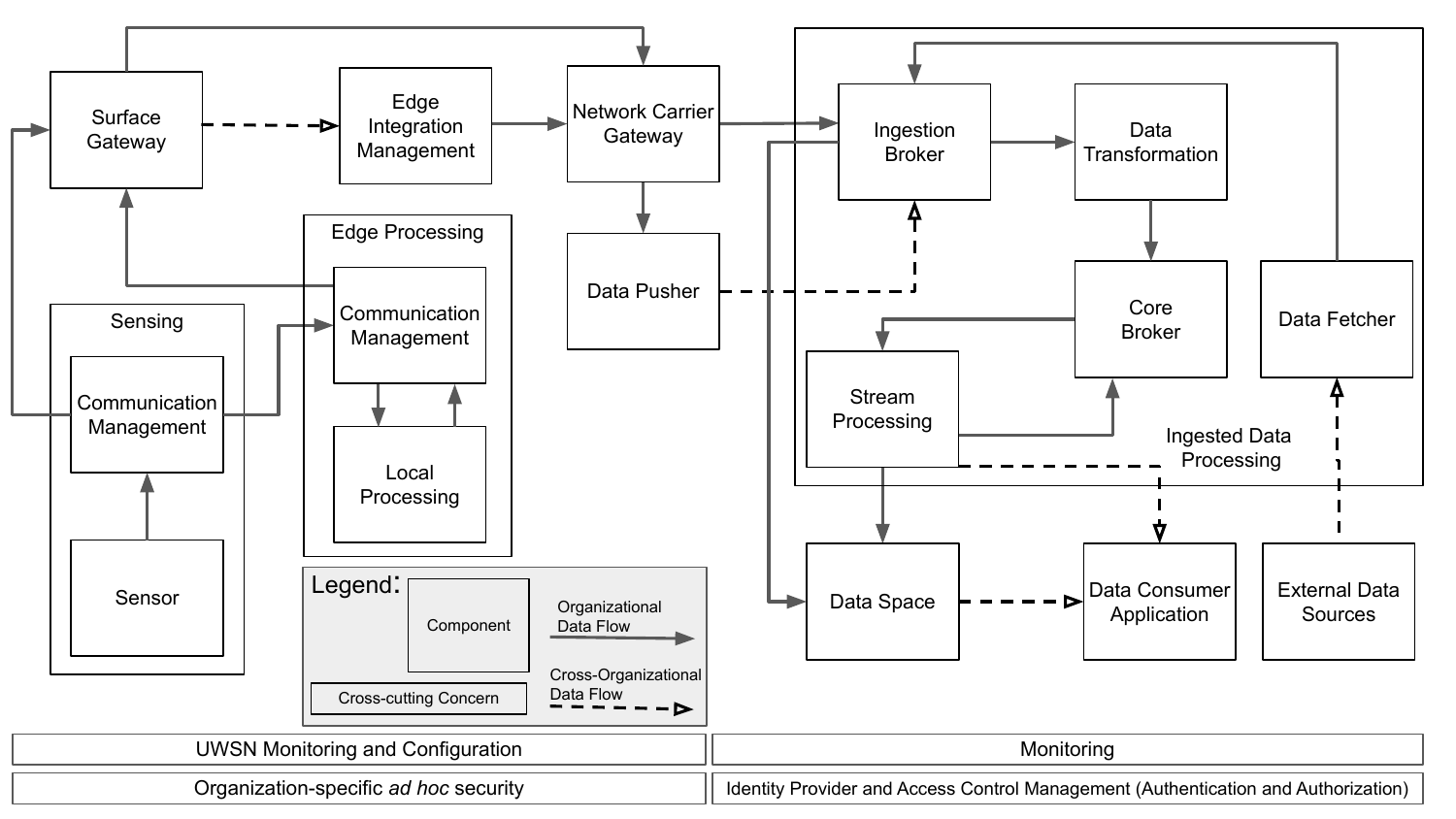}
    \caption{Overview of the resulting architecture - component and connector view.} 
    \label{fig:arch}
    \vspace{-0.3cm}
\end{figure*}

 In this section, we describe the main architectural decisions to build systems with heterogeneous marine data streams driven by data quality. The decisions are summarized in Table~\ref{tab:decisiosn}. All decisions are made by considering the challenges presented in the previous section and thus aid in mitigating the system's quality attributes affected. The architectural decisions do not address the challenges in isolation but, instead, in context with other conflicting or related challenges. 
 Throughout the description of architectural decisions, we refer also to the component and connector view shown in  Figure~\ref{fig:arch}, which highlights the data flow components.

  

\begin{table}[hb!]
\caption{Summary of Architectural Decisions (ADs)}
\label{tab:decisiosn}
\centering
\footnotesize
{
\centering
\begin{tabular}{|p{.5cm}|p{5cm}|p{2cm}|}
\hline
\textbf{ID} & \textbf{Architectural Decision} & \textbf{Challenges}\\ \hline
AD1 & Distributed data processing and transmission responsibilities & {\scriptsize CH1,CH4,CH5,CH6, CH7,CH8,CH9}  \\ \hline
AD2 & Data platform & {\scriptsize CH12} \\ \hline
AD3 & Message-oriented data ingestion middleware & {\scriptsize CH1,CH2,CH9} \\ \hline
AD4 & MQTT messaging protocol & {\scriptsize CH4,CH7,CH9}  \\ \hline
AD5 & Platform monitoring & {\scriptsize CH1,CH2,CH4, CH12,CH13,CH15}  \\ \hline
AD6 & Data quality control & {\scriptsize CH5,CH11,CH12} \\ \hline
AD7 & Data model for integrated data sources & {\scriptsize CH3,CH10,CH12} \\ \hline
AD8 & Identity provider & {\scriptsize CH13,CH15}  \\ \hline
AD9 & Data sharing access control & {\scriptsize CH13,CH14,CH15}\\ \hline
AD10 & Data classification and triage & {\scriptsize CH13,CH14}  \\ \hline
\end{tabular}
}
\end{table}

\paragraph{Distributed data processing and transmission (AD1)}
Because of the challenges at the data acquisition and network communication layers, there is a trade-off between the volume of data sent, the power consumption to transmit data (\textbf{CH9}) in the UWSN, and the density of the wireless communicating nodes in the area to cover and their impacts when transmitting data wirelessly (\textbf{CH6}). On the one hand, the volume of data depends on the number of sensor devices deployed, their configured sampling rate, security mitigation strategies (\textbf{CH8}), and the power consumption in the nodes inside the UWSN (\textbf{CH1}). These aspects are part of the responsibilities of the  \textsf{\small Sensor} component in the overall architecture layout in Figure~\ref{fig:arch}.
On the other hand, the available bandwidth and latency can vary with the water’s physical properties being in best conditions already very limited in comparision to the amount of data that can be generated or the bandwidth available in typical land-based systems, creating a bottleneck in the throughput (\textbf{CH7}). Additionally, each transmitted message inside the UWSN can impact marine life. 
The  \textsf{\small Local Processing} component in the architecture in Figure~\ref{fig:arch} is a consequence of this, being responsible for aggregating and fusing raw measurement data from the sensors in the UWSN. Because of the limitations on data transmission (\textbf{CH4} and \textbf{CH7}), this decision brings computation closer to data acquisition instead of sending raw data to be processed on the cloud level where more resources are available (\textbf{CH5}).
When it comes to the communication inside the UWSN, \textit{ad hoc}  protocols and serialization are needed when transmitting data from underwater to surface gateways. In turn, the surface gateways have to bridge to long-range transmission channels provided by satellite links or mobile networks (e.g., 4G or 5G) using the carriers' standardized APIs. This behavior is part of the responsibilities of the  \textsf{\small Communication Management} component depicted in Figure~\ref{fig:arch}, which is responsible for bridging data through heterogeneous communication channels and devices, e.g., underwater acoustics and over-the-air technologies.

\paragraph{Data Platform (AD2)} To address the interoperability issue introduced by challenge \textbf{CH12}, there was a decision to have a data platform built to allow data aggregation from different sources before its usage on the application level for use cases within the marine domain. The rationale behind this decision was to have a common configurable platform that can be parametrized when interfacing with the other layers according to its operational scenario requirements. Although some processing is done in the lowest layers, at the platform level, data fusion is performed, aggregating data from different sensing platforms. Thus, it is assumed that the volume of data is larger at this level despite sharing the property of dealing with heterogeneous data formats at the device and data source ingestion, accordingly. In the resulting architecture in Figure~\ref{fig:arch}, this decision is reflected in the  \textsf{\small Ingested Data Processing}, and the  \textsf{\small Data Space} on the right-hand side.
\paragraph{Message-oriented data ingestion middleware (AD3):} 
The rationale for this decision was to use a standardized way to provide data that is delivered by different (distributed) systems to  \textsf{\small Data Consumer Application}s. By using the publish-subscribe pattern, we decouple data-producing systems from data-consumer systems, bringing more flexibility when handling outages from each one of these types of systems. With this pattern, data producers and consumers do not need to know or keep information about each other, transferring the responsibilities to a broker component that can also handle concurrent data ingestion and provisions to the systems that might connect to it over time. The resulting architecture in Figure~\ref{fig:arch} implements this component via the  \textsf{\small Ingestion Broker}. This decision adapts to the uncertainty introduced by unreliable communications and the physical vulnerability of the components in the underwater data acquisition and the network communication layers (\textbf{CH1}, \textbf{CH2}, and \textbf{CH9}), affecting data reliability and availability.

\paragraph{MQTT Protocol (AD4)} MQTT is an OASIS standard protocol for message exchange in IoT-enabled systems and it is used on the top of the publish-subscribe communication model. Its lightweight (low footprint and small message size) characteristics and simplicity in implementation have fomented the creation of a large community and services built on top of the protocol~\cite{mqtt_survey}. Other important characteristics of the protocol are the Quality-of-Service (QoS) levels for end-to-end communication: 
Level 0 (at most once) is suitable for operational scenarios in which data loss is permitted; Level 1 (at least once) can be used in operational scenarios in which duplicated data delivery is permitted; and Level 2 (exactly once) in cases where data duplication cannot be tolerated.
Depending on the network protocol used inside the UWSN that defines the behaviour for re-transmissions for the surface gateway systems, it might be more adequate to deploy the different QoS levels from the communication between the  \textsf{\small Surface Gateway} and the  \textsf{\small Ingestion Broker} component, and addressing the uncertainty introduced by the unreliable underwater wireless communication (\textbf{CH4}).
The low footprint also fits well with the low communication bandwidth and high operation costs of the satellite link (\textbf{CH7}), and contributes to its low overhead to save bandwidth and potentially the number of transmissions needed (\textbf{CH9}).
The size of the data to be transmitted also influences 
the selection of a suitable network protocol solution.

\paragraph{Platform monitoring (AD5)}
The main drivers for this decision are the causes of faults that impact the layers inside the UWSN, affecting the availability of data on the  \textsf{\small Ingested Data Processing} component. Since there is a runtime dependency of the system behavior on data, it is crucial to detect errors that occur at runtime as early as possible to trigger context information for root cause analysis and shorten the recoverability of the overall system. More specifically, it addresses the challenges at the data acquisition and network layers,  causing uncertainty in the data availability introduced by unreliable communication, physical vulnerability, and limited battery (\textbf{CH1}, \textbf{CH2}, and \textbf{CH4}).
This decision also encompasses the monitoring of the interactions of the data platform with the external systems. 
This concerns the ingestion of data (\textbf{CH12}) and data consumption. The latter is to help assess the usage of the overall data management layer components and ensure that data is provided to rightful consumer applications systems (\textbf{CH13} and \textbf{CH15}).
For this decision to be implemented, components and data instrumentation, i.e., measurement and recording of their runtime outputted behavior, are needed. Thus, intervention in the different parts of the architecture is required making it cross-cutting concern.

\paragraph{Data Quality Control (AD6)}
Data quality is one of the main drivers and concerns of stakeholders across various organizations. It is reflected in challenge \textbf{CH11}, and understanding it for the various marine use cases is crucial for trusting the decisions made on the provided data. In this domain, because of the restricted wireless underwater communication, there is an additional level of complexity when trying to understand the data quality because of the small volume of data that can be ingested into the data management layer (\textbf{CH5}). Knowledge about the UWSN must be kept and updated to be used during the quality control process. This task is distributed across the
different processing components ( \textsf{\small Local Processing} and  \textsf{\small Stream Processing}) in Figure~\ref{fig:arch}.  In particular, metadata information about the measurement systems (sensors) is needed to understand the level of data quality and its suitability for a specific use case. Additional contextual metadata information recorded throughout the data delivery pipeline ( \textsf{\small Monitoring}) and other related historical data sources might be used to assess and ensure the quality of the data on the real-time data streams. Furthermore, the data source format standards (\textbf{CH12}) also affect the understanding of data adequacy according to the available information contained in the format in the context of different use cases.

\paragraph{Data model for integrated data sources (AD7)}   
 On one hand, as a result of the  \textsf{\small Local Processing}, there is a resulting data source in a proprietary format for each organization that is then integrated for the different  \textsf{\small Data Consumer Application}s. The integration is needed because data producers have to incorporate sensors manufactured by others and deal with different device output formats (\textbf{CH3}). On the other hand, as a consequence of the diverse data formats and standards used to ingest data from the UWSN (challenge CH12), there was a need to have a uniform format for the processing done by the different services inside the  \textsf{\small Stream Processing}. This simplifies the development of the services and addresses and fosters a single interpretation of the data at the data management layer. This decision aims to create a representative and flexible model to provide an adequate level of understanding (metadata) of data and its quality, addressing the challenges of the understandability of data for the different use cases (\textbf{CH10}). However, a challenge introduced by this decision is handling the model evolution as required information and available fields change over time and keeping backward compatibility to expected formats on the  \textsf{\small Data Consumer Application} side. Another consequence of this decision was the separation of the brokers, isolating the \textsf{\small Ingestion Broker} from  \textsf{\small Core Broker}. 

\paragraph{Identity provider (AD8)}
For the ingestion and sharing of the data in the data platform, there must be mechanisms to verify the identity of the systems interacting. There is also a need to ensure correct access, which can only be managed by knowing the identity of the systems. 
The knowledge held in the data platform intermediary layer can be used to address the access control challenge (CH15) and to establish a baseline for the selective process of sharing sensitive (CH13) or insensitive data.
During the integration of the systems in various organizations, there was also an interoperability issue among the authentication methods and solutions. The rationale behind this decision was to unify the authentication of components and their associated authorization to the different ingested and data sharing flows. Additionally, since various frameworks can be deployed between the decoupled high-level components in the architecture presented in Figure~\ref{fig:arch}, especially for the ones at the cloud level, the authentication solutions provided by those must also be integrated.
Therefore, this is a cross-cutting concern for the interactions between the components.

\paragraph{Data sharing access control (AD9)}
When data is delivered to the  \textsf{\small Data Consumer Application} component, there is a need for mechanisms to authorize access to the different flows of data (\textbf{CH15}) according to the data category and its degree of ``openness''. This decision helps to address challenges in accessing or vetting sensitive
data (\textbf{CH13}) and managing the ingestion into the  \textsf{\small Data Consumer Application}'s side because of the reusability challenge (\textbf{CH14}). This decision relies on an identity solution for the interacting systems and is thus closely related to AD8. In conjunction, they constitute a cross-cutting concern regarding ensuring security at the data management layer is represented by \textsf{\small Identity Provider and Access Control Management}  in Figure~\ref{fig:arch}. Its implementation is based on adding authentication to the different components and sub-components, and by adding authorization to data accordingly. This decision encompasses rules defined in authentication and authorization management, integrating with the various ways to deliver data via the  \textsf{\small Core Broker} and the  \textsf{\small Data Space} using the push and pull methods to the  \textsf{\small Data Consumer Application}. Moreover, this might also entail granularity of controls at brokers' topics access.

\paragraph{Data classification and triage (AD10)}
Complementary to the access control decision (AD9), an intermediary step is needed to classify the ingested data streams. One of the major drivers behind this decision is the sensitive data (\textbf{CH13}) because of its reusability (\textbf{CH14}) in a context different from the one for which the data acquisition was designed. The rationale for this decision is to filter different categories, quality levels, and ``openness'' of data impacting the access control to data shared on the  \textsf{\small Data Consumer Application} components. In the resulting architecture in Figure~\ref{fig:arch}, the  \textsf{\small Stream Processing} component is responsible for implementing the behavior behind the rationale of this decision.

\section{Architecture in Context} 
\label{sec:prototype_arch}
\begin{figure}[htb!] 
	\centering 
	\begin{subfigure}{\linewidth}
    	\centering 
    	\includegraphics[trim={5.2cm 0.3cm 0cm 0cm},clip,width=\linewidth]{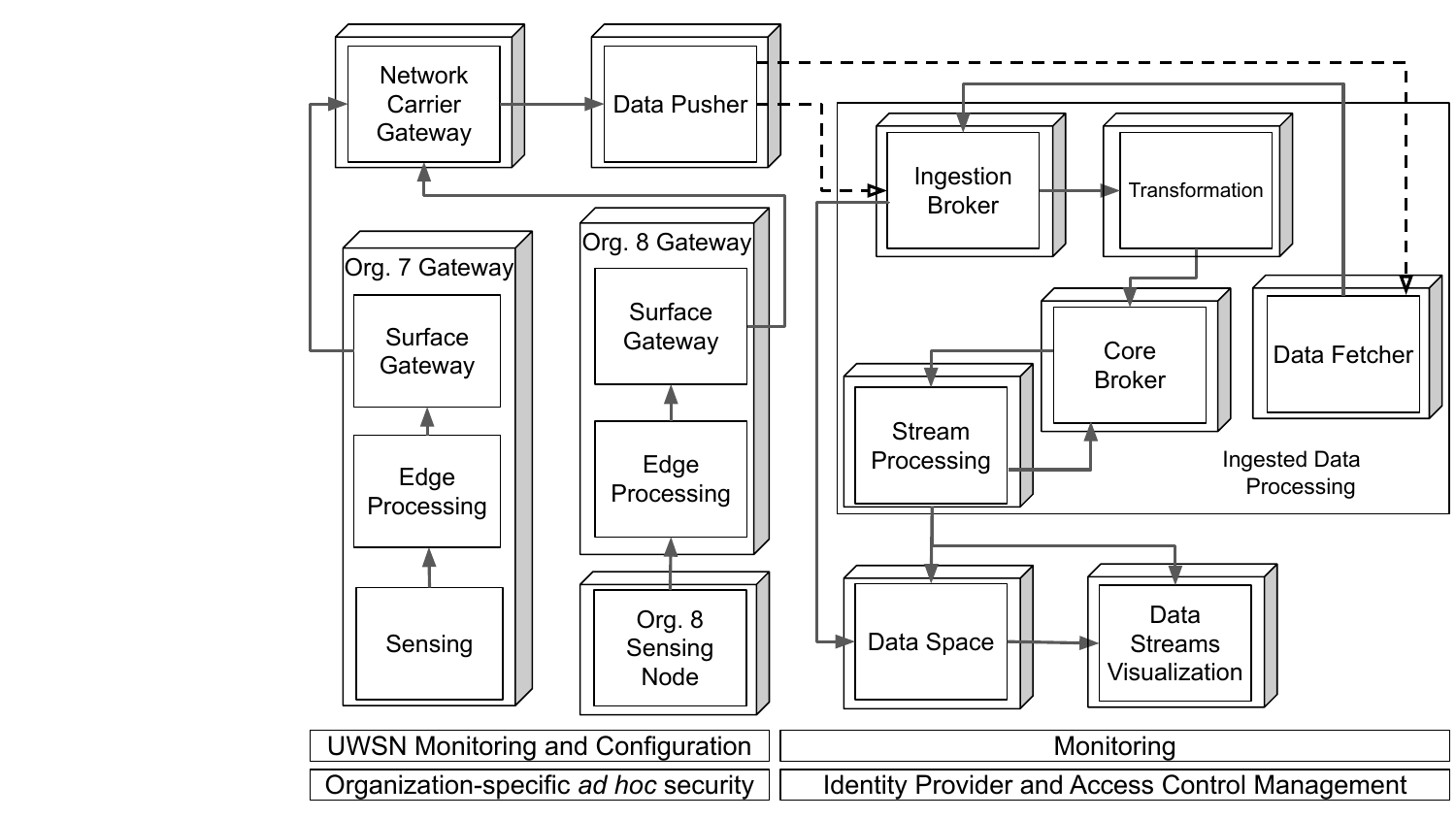}
	\end{subfigure}
    \hfill
    \begin{subfigure}{\linewidth}
    	\centering 
	\includegraphics[trim={5.1cm 1.2cm 0cm 10.3cm},clip,width=\linewidth]{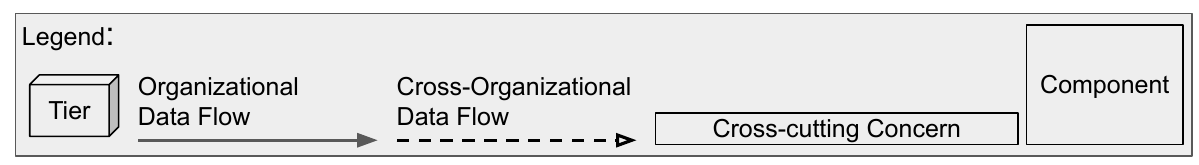}
	\end{subfigure}
    \caption{Architecture of the prototype implementation.}
    \label{fig:prototype}
\end{figure}

The prototype implementation presented in Figure~\ref{fig:prototype} reflects the ongoing migration process to the proposed architecture. 
To realize this migration, many factors come into play, making it a complex and slow process. More specifically, the cross-organizational dependency of the development and operation of different component blocks. 

The tiers, graphically shown as a cube in Figure~\ref{fig:prototype}'s legend (\textsf{\small Tier} label), represent physical machines at the UWSN level while at the cloud level, the components are allocated to hybrid clouds using private and public cloud providers.

In the prototype, there are currently two data producers organizations with different data acquisition and network communication setups, as we can see in Figure~\ref{fig:prototype}.
Organization 7 has in its portfolio a vertically integrated sensor (physical) platform that incorporates all the needed components for the data acquisition and network layers. There are currently two of these platforms deployed at sea and sending data approximately every 30 minutes to their control and storage platform.
While this setup avoids the use of wireless underwater communication, it also presents a single point of failure and needs to be combined with the deployment of other platforms to expand the area coverage.
Oganization 8 has a more flexible setup, with a mesh of nodes distributed in the same area of interest and with adaptive roles locally coordinated by their gateway at the edge of the UWSN. There are currently two sensing nodes and one gateway sending data to their data platform approximately every hour.
Both organizations are using mobile networks to transmit data over-the-air to their data platform, using the  \textsf{\small Network Carrier Gateway} component.

For the security inside the UWSN, because of the flexibility and distribution of components, there might be no wireless underwater data transmission and thus, the challenge \textbf{CH9} (see Table~\ref{tab:challenges}) associated with that link does not apply. This is the case for organization's 7 system's components architecture. It employs all the needed components in the same physical system with serial or ethernet connections from the sensors to the surface gateway. On the other hand,  there could also be a mesh of nodes physically distributed, which is the case for the configuration of the system for organization 8. The security inside their UWSN is provided by a tailored framework that follows the NIST recommended levels of security in~\cite{nist_csf} with focus on low overhead in the underwater communications.

At the cloud level, the authors associated organization is responsible for the implementation of the components at the data management layer for the heterogeneous data streams ingested by the data producers' systems. The implemented  \textsf{\small Ingested Data Processing} component is currently transforming, quality controlling, and forwarding data sources ingested in JSON and XML formats by the different data producers. The data quality control checks whether a recently data point is good or bad. The check is triggered once a data point arrives to cope with real-time data usage requirement from data consumers. In case a data point is missing due to, for example, communication issues, sensor faults, the component generates a missing value and triggers an alarm. The quality of data is assessed by four data quality attributes as introduced in Section~\ref{sec:intrinsics_QA}. 
The cross-organizational integration done at the cloud level and the use of message-oriented middleware for data ingestion (AD3), resulted the development of a new way of delivering data for the existing data platform of organizations 7 and 8 (see Table~\ref{tab:partners}). These changes were  done to implement the expected behavior for the  \textsf{\small Data Pusher} component of the overall architecture in Figure~\ref{fig:arch}, and receiving the data as soon as it arrives on the different clouds. For organization 7 all the efforts where done on their side to complete the migration process. For organization 8 data integration using the push behavior, the  \textsf{\small Data Pusher} component was implemented on the data management side to bridge the data provided via their own MQTT broker. For the broker components implementation on the prototype we are using HiveMQ (\url{https://www.hivemq.com/}).
Lastly, regarding data ingestion, since the data was initially provided by organization 8 via a REST API using the  \textsf{\small Data Fetcher} component, this ingestion method can still be used to retrieve historical data in case of down time (availability issues) on the sending or receiving components. 

For the  \textsf{\small Monitoring} component, we are monitoring Key Performance Indicators (KPI) for the data platform. These KPIs function as service level objectives and are translated to interact with the Prometheus  monitoring toolkit (\url{https://prometheus.io/}) where runtime metrics are stored.

Concerning the data provision to  \textsf{\small Data Consumer Application}s, the current prototype is dealing with heterogeneous authentication methods challenge being in the process of migrating to a centralized authentication method using Keycloak (\url{https://www.keycloak.org/})  as an identity and access management solution to enable the access control via the  services' identity associated credentials. 
Nevertheless, for testing proposes, a  \textsf{\small Data Streams Visualization} GUI has being integrated to visualize the data coming from the different streams. This component was implemented using Thingspeak (\url{https://thingspeak.com}).
Finally, on the  \textsf{\small Data Consumer Application} side, efforts are being made to deliver data to a data visualization and analytics software operated by organization 2, and to a marine data archive managed by organization 13 (see Table~\ref{tab:partners}).  

\section{Discussion} 
\label{sec:discussion}
The architectural decisions proposed conceptually address concerns raised by the identified challenges, providing a clear understanding of data quality in the different stages of the data flow: data acquisition, processing, and (re)usage. 
At the Underwater Wireless Sensor Network (UWSN) level, the components are managed by the organization that developed and deployed them as a service. 
Because of the custom hardware setup of the data producers’ systems, the data can flow among the processing and transmitting components, traversing different traces according to the routing protocols inside the UWSN. From the data producer actor viewpoint, the components at this level must optimize the data to be sent up, prioritizing what is more relevant for the operations in case there are conflicts in the bandwidth usage. At the same time, there is a need to balance the battery consumption, the environmental impact of the data delivery in (near) real-time, and the network security measures (\textbf{AD1}). 

Regarding the cross-cutting concerns at the UWSN level, on the one hand, the data producer systems can monitor and configure the different nodes remotely. 
On the other hand, concerning security, there are no domain-specific solutions to address the challenge of securing underwater wireless communications in the case of obstruction of the communication channels (denial-of-service) or a compromised node (spoofing) since the wireless communication channels underwater are open. 
The systems’ resource capacity imposes constraints on adopting solutions from other wireless communication and IP networks. Overall, these requirements enable the reliable and secure adjustment of nodes’ behavior regarding the volume of data needed for acquisition and transmission.


At the cloud level, the  \textsf{\small Ingested Data Processing} sub-components are loosely coupled for the elasticity of the computation due to the variable volume of data pushed or pulled into the pipeline. Additionally, data might need different processing services according to the quality requirements. 
The first step into the processing of heterogeneous data streams ingested is to convert them to a complying syntax and semantics, assuring that all the processing components work upon the same specification and understanding of data (\textbf{AD7}). This step is also responsible for validating the ingested data, contributing to the completeness and compliance quality attributes of data. If the storage of data in its raw format is desired for future reprocessing, this can be achieved by the proposed architecture. 
After the transformation process, data is again ingested into the \textsf{\small Core
Broker} component for distribution to the different  \textsf{\small Stream Processing} services. Additional information might be needed by these processing services not only at the UWSN level but also from external data. Another responsibility held by the  \textsf{\small Stream Processing} component is data categorization, aiding the sharing access control rules implementation, and addressing availability and confidentiality data quality attributes.
The data model will have significant impact on the selection of solutions for the database management system (\textsf{\small Data Space} component in Figure~\ref{fig:arch}). It is important to have a flexible database model to minimize the overload when updating the components with the evolution of the data and metadata fields requirements.

Regarding the cross-organizational data flow, to enable the flexibility on the delivery of data needed at the cloud side, there are three alternatives for data ingestion in the data management layer. In the proposed architecture, this can be achieved via the pusher, fetcher or the edge components.
The first component is responsible to integrate data sources at the cloud level interacting with data producers' systems over IP networks. In this case of data ingestion, the responsibility lays on the data producers side to push data into the data management layer.
Second, the  \textsf{\small Data Fetcher} component can pull data from ephemeral data acquisition systems (manual/ship sea surveys), other databases or streams into the processing pipeline.
Thirdly, the \textsf{\small Edge Integration Management} component differs from the others in its physical location at the UWSN level. In this case, the data sources integration is performed closer to the data acquisition, with more access to information in this layer, transferring the costs of OTA transmissions done via the  \textsf{\small Network Carrier Gateway} component. For this component to be implemented, the data producers' systems have to be exposed for integration locally.
The existence of neither one of these three components is exclusive, this is a reflection of the distribution of processing and communication responsibilities in the first architectural decision (see Table~\ref{tab:decisiosn}), as was demonstrated in the architecture implementation in Figure~\ref{fig:prototype}.

Finally, on the cross-organizational data flow, there is a need to make heterogeneous components' authentication methods uniform to enforce ingestion access policies and sharing rules. There is also a need to regulate the access to data via push and pull methods (from the  \textsf{\small Stream Processing} or the  \textsf{\small Data Space} respectively). This is similar to the ingestion into the data management layer. These concerns are transversal to the architecture and must be implemented in the organizational boundaries along the data flow.
%

\section{Related Work}
\label{sec:related_work}

Important aspects of data architecture for Internet-of-Things 
(IoT) applications are~\cite{abughazala2022dat}: data quality, data security, organizational efficiency, business intelligence and analytics, and scalability and flexibility. IoT is the foundation of IoUT in which \texttt{System} is situated~\cite{pena2005itu}. We took into account those aspects while designing our architecture.

Some software architectures have been proposed to accommodate heterogeneous aspects~\cite{seifermann2019data,rademacher2019aspect,CORRALPLAZA2020103426}. However, none of them considers a broad range of software and data quality concerns as we have identified, and neither consider a data actor viewpoint. The work in \cite{seifermann2019data} deals with the confidentiality of multiple data streams. A metamodel built on top of the Architectural Description Language (ADL) Palladio Component Model (PCM)~\cite{reussner2016modeling} is proposed to detect access right mismatches regarding data consumption roles of an information system. This solution covers only two challenges (CH13 and CH15 in Table~\ref{tab:challenges}). In IoT environments, the work in ~\cite{CORRALPLAZA2020103426} proposes an architecture focusing on the heterogeneous data processing layer based on nine requirements from 2 survey studies, abstracting the data sources and consumers.
The paper ~\cite{rademacher2019aspect} addresses the challenges of accommodating various different technologies used in information systems. There are different software technologies utilized in our software architecture (see Section~\ref{sec:decisions}). We rely on Service-oriented Architecture (SOA)~\cite{siqueira2021service, hastbacka2019dynamic} to decouple the components and let stakeholders take responsibility of maintaining their components independently. We leverage MQTT as a plug and play mechanism~\cite{hastbacka2019dynamic} to facilitate migrations of various data sources to the system and address the challenge of having multiple communication protocols, data formats, and data streams in \texttt{System}.

More recently, an architecture for an IoUT system has been proposed based on four multi-vocal references and validated with three quantitative performance efficiency metrics~\cite{RAZZAQ2023100893}.
Our work has a concrete context enhanced by challenges gathered from nineteen studies from the literature. Consequently, our validation has a broader scope rather than performance quality attributes metrics, allowing us to take into account factors affecting the quality of not only software (like~\cite{ref13}) but also data (see~\cite{ref15}) while designing the architecture.

\section{Conclusion and Future Work}
\label{sec:conclusion}
This study employed the analysis and discussion of quality attributes for software and data impacted by domain-specific challenges in a cross-organizational context. We presented a set of architectural decisions that incorporate the concerns from the challenges. The resulting data-flow-oriented architecture has been instantiated and demonstrated in a prototype implementation, which has been built in the context of a research and development project. As more iterations occur, we expect new challenges, especially concerning data sharing for reusability purposes, edge integration, and backward flow of control if exposed as a cross-organizational service.
In future works, the relation between data and software quality attributes will be further investigated for the evaluation of the architecture, including how they can be quantified and measured (metrics). 
In particular, the credibility data quality attribute needs to be further investigated as it remains an open challenge and goes beyond the scope of this paper.

\section*{Acknowledgment}
This work was partly supported by SFI SmartOcean NFR Project 309612/F40, and  PNRR MUR project VITALITY (ECS00000041), Spoke 2 ASTRA - ``Advanced Space Technologies and Research Alliance"
and of the MUR (Italy) Department of Excellence 2023 - 2027 for GSSI. 

\bibliographystyle{IEEEtran}
\bibliography{main}

\begin{thebibliography}{10}
\providecommand{\url}[1]{#1}
\csname url@samestyle\endcsname
\providecommand{\newblock}{\relax}
\providecommand{\bibinfo}[2]{#2}
\providecommand{\BIBentrySTDinterwordspacing}{\spaceskip=0pt\relax}
\providecommand{\BIBentryALTinterwordstretchfactor}{4}
\providecommand{\BIBentryALTinterwordspacing}{\spaceskip=\fontdimen2\font plus
\BIBentryALTinterwordstretchfactor\fontdimen3\font minus \fontdimen4\font\relax}
\providecommand{\BIBforeignlanguage}[2]{{%
\expandafter\ifx\csname l@#1\endcsname\relax
\typeout{** WARNING: IEEEtran.bst: No hyphenation pattern has been}%
\typeout{** loaded for the language `#1'. Using the pattern for}%
\typeout{** the default language instead.}%
\else
\language=\csname l@#1\endcsname
\fi
#2}}
\providecommand{\BIBdecl}{\relax}
\BIBdecl

\bibitem{ref14}
N.-T. Nguyen, R.~Heldal, K.~Lima, T.~D. Oyetoyan, P.~Pelliccione, L.~M. Kristensen, K.~W. Høydal, P.~A. Reiersgaard, and Y.~Kvinnsland, ``Engineering challenges of stationary wireless smart ocean observation systems,'' \emph{IEEE Internet of Things Journal}, vol.~10, no.~16, pp. 14\,712--14\,724, 2023.

\bibitem{ref13}
K.~Lima, N.-T. Nguyen, R.~Heldal, E.~Knauss, T.~D. Oyetoyan, P.~Pelliccione, and L.~M. Kristensen, ``Marine data sharing: Challenges, technology drivers and quality attributes,'' in \emph{Product-Focused Software Process Improvement}, D.~Taibi, M.~Kuhrmann, T.~Mikkonen, J.~Kl{\"u}nder, and P.~Abrahamsson, Eds.\hskip 1em plus 0.5em minus 0.4em\relax Cham: Springer International Publishing, 2022, pp. 124--140.

\bibitem{10188681}
V.~Chamola, V.~Hassija, A.~R. Sulthana, D.~Ghosh, D.~Dhingra, and B.~Sikdar, ``A review of trustworthy and explainable artificial intelligence (xai),'' \emph{IEEE Access}, vol.~11, pp. 78\,994--79\,015, 2023.

\bibitem{9609199}
G.~A. Lewis, I.~Ozkaya, and X.~Xu, ``Software architecture challenges for ml systems,'' in \emph{2021 IEEE International Conference on Software Maintenance and Evolution (ICSME)}, 2021, pp. 634--638.

\bibitem{kuwajima2020engineering}
H.~Kuwajima, H.~Yasuoka, and T.~Nakae, ``Engineering problems in machine learning systems,'' \emph{Machine Learning}, vol. 109, no.~5, pp. 1103--1126, 2020.

\bibitem{hevner2010design}
A.~Hevner and S.~Chatterjee, \emph{Design research in information systems: theory and practice}.\hskip 1em plus 0.5em minus 0.4em\relax Springer Science \& Business Media, 2010, vol.~22.

\bibitem{petersen2009context}
K.~Petersen and C.~Wohlin, ``Context in industrial software engineering research,'' in \emph{2009 3rd International Symposium on Empirical Software Engineering and Measurement}.\hskip 1em plus 0.5em minus 0.4em\relax IEEE, 2009, pp. 401--404.

\bibitem{ref15}
N.-T. Nguyen, K.~Lima, A.~M. Skålvik, R.~Heldal, E.~Knauss, T.~D. Oyetoyan, P.~Pelliccione, and C.~Sætre, ``Synthesized data quality requirements and roadmap for improving reusability of in-situ marine data,'' in \emph{2023 IEEE 31st International Requirements Engineering Conference (RE)}, 2023, pp. 65--76.

\bibitem{replication_package}
\BIBentryALTinterwordspacing
K.~Lima, ``{A data-flow oriented software architecture for heterogeneous marine data streams: Appendix},'' Feb. 2024. [Online]. Available: \url{https://doi.org/10.5281/zenodo.10653178}
\BIBentrySTDinterwordspacing

\bibitem{iso25010}
J.~Estdale and E.~Georgiadou, ``Applying the iso/iec 25010 quality models to software product,'' in \emph{Systems, Software and Services Process Improvement}, X.~Larrucea, I.~Santamaria, R.~V. O'Connor, and R.~Messnarz, Eds.\hskip 1em plus 0.5em minus 0.4em\relax Cham: Springer International Publishing, 2018, pp. 492--503.

\bibitem{iso25012_2}
\BIBentryALTinterwordspacing
F.~Gualo, M.~Rodriguez, J.~Verdugo, I.~Caballero, and M.~Piattini, ``Data quality certification using iso/iec 25012: Industrial experiences,'' \emph{Journal of Systems and Software}, vol. 176, p. 110938, 2021. [Online]. Available: \url{https://www.sciencedirect.com/science/article/pii/S0164121221000352}
\BIBentrySTDinterwordspacing

\bibitem{bass2003software}
L.~Bass, P.~Clements, and R.~Kazman, \emph{Software architecture in practice}.\hskip 1em plus 0.5em minus 0.4em\relax Addison-Wesley Professional, 2003.

\bibitem{habibullah2023non}
K.~M. Habibullah, G.~Gay, and J.~Horkoff, ``Non-functional requirements for machine learning: Understanding current use and challenges among practitioners,'' \emph{Requirements Engineering}, vol.~28, no.~2, pp. 283--316, 2023.

\bibitem{VENTERS2018174}
\BIBentryALTinterwordspacing
C.~C. Venters, R.~Capilla, S.~Betz, B.~Penzenstadler, T.~Crick, S.~Crouch, E.~Y. Nakagawa, C.~Becker, and C.~Carrillo, ``Software sustainability: Research and practice from a software architecture viewpoint,'' \emph{Journal of Systems and Software}, vol. 138, pp. 174--188, 2018. [Online]. Available: \url{https://www.sciencedirect.com/science/article/pii/S0164121217303072}
\BIBentrySTDinterwordspacing

\bibitem{koccak2015integrating}
S.~A. Ko{\c{c}}ak, G.~I. Alptekin, and A.~B. Bener, ``Integrating environmental sustainability in software product quality.'' in \emph{RE4SuSy@ RE}, 2015, pp. 17--24.

\bibitem{ref2}
\BIBentryALTinterwordspacing
F.~Audo, M.~Guegan, V.~Quintard, A.~Perennou, J.~L. Bihan, and Y.~Auffret, ``{Quasi-all-optical network extension for submarine cabled observatories},'' \emph{Optical Engineering}, vol.~50, no.~4, p. 045001, 2011. [Online]. Available: \url{https://doi.org/10.1117/1.3560542}
\BIBentrySTDinterwordspacing

\bibitem{ref3}
\BIBentryALTinterwordspacing
M.~C. Domingo, ``An overview of the internet of underwater things,'' \emph{Journal of Network and Computer Applications}, vol.~35, no.~6, pp. 1879--1890, 2012. [Online]. Available: \url{https://www.sciencedirect.com/science/article/pii/S1084804512001646}
\BIBentrySTDinterwordspacing

\bibitem{ref5}
T.~Qiu, Z.~Zhao, T.~Zhang, C.~Chen, and C.~L.~P. Chen, ``Underwater internet of things in smart ocean: System architecture and open issues,'' \emph{IEEE Transactions on Industrial Informatics}, vol.~16, no.~7, pp. 4297--4307, 2020.

\bibitem{ref11}
M.~Jahanbakht, W.~Xiang, L.~Hanzo, and M.~Rahimi~Azghadi, ``Internet of underwater things and big marine data analytics—a comprehensive survey,'' \emph{IEEE Communications Surveys \& Tutorials}, vol.~23, no.~2, pp. 904--956, 2021.

\bibitem{ref4}
\BIBentryALTinterwordspacing
C.-C. Kao, Y.-S. Lin, G.-D. Wu, and C.-J. Huang, ``A comprehensive study on the internet of underwater things: Applications, challenges, and channel models,'' \emph{Sensors}, vol.~17, no.~7, 2017. [Online]. Available: \url{https://www.mdpi.com/1424-8220/17/7/1477}
\BIBentrySTDinterwordspacing

\bibitem{ref7}
\BIBentryALTinterwordspacing
S.~Fattah, A.~Gani, I.~Ahmedy, M.~Y.~I. Idris, and I.~A. Targio~Hashem, ``A survey on underwater wireless sensor networks: Requirements, taxonomy, recent advances, and open research challenges,'' \emph{Sensors}, vol.~20, no.~18, 2020. [Online]. Available: \url{https://www.mdpi.com/1424-8220/20/18/5393}
\BIBentrySTDinterwordspacing

\bibitem{ref8}
M.~Han, J.~Duan, S.~Khairy, and L.~X. Cai, ``Enabling sustainable underwater iot networks with energy harvesting: A decentralized reinforcement learning approach,'' \emph{IEEE Internet of Things Journal}, vol.~7, no.~10, pp. 9953--9964, 2020.

\bibitem{ref9}
M.~F. Ali, D.~N.~K. Jayakody, Y.~A. Chursin, S.~Affes, and S.~Dmitry, ``Recent advances and future directions on underwater wireless communications,'' \emph{Archives of Computational Methods in Engineering}, vol.~27, pp. 1379--1412, 2020.

\bibitem{ref10}
\BIBentryALTinterwordspacing
P.~Mariani, R.~Bachmayer, S.~Kosta, E.~Pietrosemoli, M.~V. Ardelan, D.~P. Connelly, E.~Delory, J.~S. Pearlman, G.~Petihakis, F.~Thompson, and A.~Crise, ``Collaborative automation and iot technologies for coastal ocean observing systems,'' \emph{Frontiers in Marine Science}, vol.~8, 2021. [Online]. Available: \url{https://www.frontiersin.org/articles/10.3389/fmars.2021.647368}
\BIBentrySTDinterwordspacing

\bibitem{ref1}
M.~Stojanovic and J.~Preisig, ``Underwater acoustic communication channels: Propagation models and statistical characterization,'' \emph{IEEE Communications Magazine}, vol.~47, no.~1, pp. 84--89, 2009.

\bibitem{ref6}
S.~Aslam, M.~P. Michaelides, and H.~Herodotou, ``Internet of ships: A survey on architectures, emerging applications, and challenges,'' \emph{IEEE Internet of Things Journal}, vol.~7, no.~10, pp. 9714--9727, 2020.

\bibitem{ref16}
A.~Zolich, D.~Palma, K.~Kansanen, K.~Fj{\o}rtoft, J.~Sousa, K.~H. Johansson, Y.~Jiang, H.~Dong, and T.~A. Johansen, ``Survey on communication and networks for autonomous marine systems,'' \emph{Journal of Intelligent \& Robotic Systems}, vol.~95, pp. 789--813, 2019.

\bibitem{ref12}
\BIBentryALTinterwordspacing
D.~R.~K. Mary, E.~Ko, S.-G. Kim, S.-H. Yum, S.-Y. Shin, and S.-H. Park, ``A systematic review on recent trends, challenges, privacy and security issues of underwater internet of things,'' \emph{Sensors}, vol.~21, no.~24, 2021. [Online]. Available: \url{https://www.mdpi.com/1424-8220/21/24/8262}
\BIBentrySTDinterwordspacing

\bibitem{ref19}
\BIBentryALTinterwordspacing
A.~G. Yisa, T.~Dargahi, S.~Belguith, and M.~Hammoudeh, ``Security challenges of internet of underwater things: A systematic literature review,'' \emph{Transactions on Emerging Telecommunications Technologies}, vol.~32, no.~3, p. e4203, 2021. [Online]. Available: \url{https://onlinelibrary.wiley.com/doi/abs/10.1002/ett.4203}
\BIBentrySTDinterwordspacing

\bibitem{ref17}
Z.~Tan, B.~Zhang, X.~Wu, M.~Dong, and L.~Cheng, ``Quality control for ocean observations: From present to future,'' \emph{Science China Earth Sciences}, pp. 1--18, 2022.

\bibitem{ref18}
J.~J. Buck, S.~J. Bainbridge, E.~F. Burger, A.~C. Kraberg, M.~Casari, K.~S. Casey, L.~Darroch, J.~D. Rio, K.~Metfies, E.~Delory \emph{et~al.}, ``Ocean data product integration through innovation-the next level of data interoperability,'' \emph{Frontiers in Marine Science}, vol.~6, p.~32, 2019.

\bibitem{iso25012}
\BIBentryALTinterwordspacing
H.~Foidl and M.~Felderer, ``An approach for assessing industrial iot data sources to determine their data trustworthiness,'' \emph{Internet of Things}, vol.~22, p. 100735, 2023. [Online]. Available: \url{https://www.sciencedirect.com/science/article/pii/S2542660523000586}
\BIBentrySTDinterwordspacing

\bibitem{mqtt_survey}
B.~Mishra and A.~Kertesz, ``The use of mqtt in m2m and iot systems: A survey,'' \emph{IEEE Access}, vol.~8, pp. 201\,071--201\,086, 2020.

\bibitem{nist_csf}
{National Institute of Standards and Technology}, ``Public draft: The nist cybersecurity framework 2.0,'' 2023.

\bibitem{abughazala2022dat}
M.~Abughazala, H.~Muccini, and M.~Sharaf, ``Dat: Data architecture modeling tool for data-driven applications,'' in \emph{European Conference on Software Architecture}.\hskip 1em plus 0.5em minus 0.4em\relax Springer, 2022, pp. 90--101.

\bibitem{pena2005itu}
I.~Pe{\~n}a-L{\'o}pez \emph{et~al.}, ``Itu internet report 2005: the internet of things,'' 2005.

\bibitem{seifermann2019data}
S.~Seifermann, R.~Heinrich, and R.~Reussner, ``Data-driven software architecture for analyzing confidentiality,'' in \emph{2019 IEEE International Conference on Software Architecture (ICSA)}.\hskip 1em plus 0.5em minus 0.4em\relax IEEE, 2019, pp. 1--10.

\bibitem{rademacher2019aspect}
F.~Rademacher, S.~Sachweh, and A.~Z{\"u}ndorf, ``Aspect-oriented modeling of technology heterogeneity in microservice architecture,'' in \emph{2019 IEEE International conference on software architecture (ICSA)}.\hskip 1em plus 0.5em minus 0.4em\relax IEEE, 2019, pp. 21--30.

\bibitem{CORRALPLAZA2020103426}
\BIBentryALTinterwordspacing
D.~Corral-Plaza, I.~Medina-Bulo, G.~Ortiz, and J.~Boubeta-Puig, ``A stream processing architecture for heterogeneous data sources in the internet of things,'' \emph{Computer Standards \& Interfaces}, vol.~70, p. 103426, 2020. [Online]. Available: \url{https://www.sciencedirect.com/science/article/pii/S092054891930008X}
\BIBentrySTDinterwordspacing

\bibitem{reussner2016modeling}
R.~H. Reussner, S.~Becker, J.~Happe, R.~Heinrich, and A.~Koziolek, \emph{Modeling and simulating software architectures: The Palladio approach}.\hskip 1em plus 0.5em minus 0.4em\relax MIT Press, 2016.

\bibitem{siqueira2021service}
F.~Siqueira and J.~G. Davis, ``Service computing for industry 4.0: State of the art, challenges, and research opportunities,'' \emph{ACM Computing Surveys (CSUR)}, vol.~54, no.~9, pp. 1--38, 2021.

\bibitem{hastbacka2019dynamic}
D.~H{\"a}stbacka, J.~Halme, M.~Larra{\~n}aga, R.~More, H.~Mesi{\"a}, M.~Bj{\"o}rkbom, L.~Barna, H.~Pettinen, M.~Elo, A.~Jaatinen \emph{et~al.}, ``Dynamic and flexible data acquisition and data analytics system software architecture,'' in \emph{2019 IEEE SENSORS}.\hskip 1em plus 0.5em minus 0.4em\relax IEEE, 2019, pp. 1--4.

\bibitem{RAZZAQ2023100893}
\BIBentryALTinterwordspacing
A.~Razzaq, A.~Ahmad, A.~W. Malik, M.~Fahmideh, and R.~A. Ramadan, ``Software engineering for internet of underwater things to analyze oceanic data,'' \emph{Internet of Things}, vol.~24, p. 100893, 2023. [Online]. Available: \url{https://www.sciencedirect.com/science/article/pii/S2542660523002160}
\BIBentrySTDinterwordspacing

\end{thebibliography}

\end{document}